\documentclass[smallextended]{svjour3}       
\smartqed  
\usepackage{graphicx}
\begin{document}

\title{New constraints on $H_0$ and $\Omega_m$ from SZE/X-RAY data and
Baryon Acoustic Oscillations
}


\author{R. F. L. Holanda        \and
        J. V. Cunha   \and     J. A. S. Lima 
}


\institute{R. F. L. Holanda \and J. V. cunha \and J. A. S. Lima 
\at Departamento de Astronomia, Universidade de S\~ao Paulo, Rua do Mat\~ao, 1226 - 05508-900, S\~ao Paulo, SP, Brazil\\
,\\\email{holanda@astro.iag.usp.br}\\
\email{cunhajv@astro.iag.usp.br}\\
\email{limajas@astro.iag.usp.br}
}
\date{Received: date / Accepted: date}

\maketitle

\begin{abstract}
The Hubble constant, $H_0$, sets the scale of the size and age of the Universe
and its determination from independent methods is still worthwhile to be investigated.
In this article,  by using the Sunyaev-Zeldovich effect and X-ray surface brightness
data from 38 galaxy clusters observed by Bonamente {\it{et al.}} (2006), we obtain
a new estimate of $H_0$ in the context of  a flat $\Lambda$CDM model.
There is a degeneracy on the mass density parameter ($\Omega_{m}$)
which is broken by applying a joint analysis involving the baryon acoustic
oscillations (BAO) as given by Sloan Digital Sky Survey (SDSS). This
happens because the BAO signature does not depend on $H_0$. Our basic finding
is that a joint analysis involving these tests yield $H_0=
0.765^{+0.035}_{-0.033}$ km s$^{-1}$ Mpc$^{-1}$
and $\Omega_{m}=0.27^{+0.03}_{-0.02}$. Since the hypothesis of spherical
geometry assumed by Bonamente  {\it {et al.}} is questionable,
we have also compared the above results to a recent work
where a sample of galaxy clusters  described by an elliptical profile was used in analysis.

\end{abstract}

\section{Introduction}
\label{intro}

The so-called Sunyaev-Zeldovich effect (SZE) is a small distortion on the Cosmic Microwave
Background (CMB) spectrum provoked by the inverse Compton scattering
of the CMB photons passing through a population of hot
electrons, such as intra galaxy cluster medium \cite{SunZel72,Itoh98}. The SZE is proportional to the pressure intregrated along the line of sight, $\Delta T_{SZE}/T_0 \propto \int n_{e} T_{e}dl$, where $T_0$ is the CMB temperature, $n_e$ and $T_e$ are the density and temperature of electrons in intracluster gas, and its magnitude is $\Delta T_{SZE}/T_0 \approx 10^{-5}$ independent of redshift. On the other hand, X-ray emission from the intracluster gas has a different dependence on the density of electrons, $S_{X}\propto \int n_{e}^{2}\Lambda_{eH}$, where $\Lambda_{eH}$ is the X-ray cooling function. It is possible to take advantage of the different density dependencies in these phenomena and evaluate the angular diameter distance (ADD) of the galaxy cluster by \cite{Barttlet04,SilWhi78,Caval79}
\begin{equation}
D_{A}(z)\propto \frac{(\Delta T_{0})^{2}\Lambda_{eH0}}{S_{X0}T_{e0}^{2}}\frac{(1+z)^{-4}}{\theta_{c}},
\end{equation}
{where $z$ is the redshift os galaxy cluster and $\theta_{c}$ is its  core radius obtained from the SZE and X-ray analysis}. Hence, it is possible to construct a Hubble diagram with which some cosmological parameters, such as, matter density, dark energy density, Hubble parameter ($H_{0}$), etc, can be estimated \cite{Barttlet04}.

This technique for measuring distances is completely independent of other methods (as the one provided by the luminosity distance), and it can be used to measure distances at high redshifts directly. Review
papers on this subject have been published by Birkinshaw\cite{Birki99} and Carlstrom, Holder and
Reese\cite{Carlstrom02}. More recently,  such a technique has been applied for a fairly large number of
clusters\cite{Reese02,Jones05,DeFilippis05,Boname06,Roque06,Mas01}.

However, in order to estimate the ADD of a galaxy cluster from its SZE/X-ray observations, one needs to add some complementary assumptions about its geometry. In the last decade, many studies about the intracluster gas and dark
matter distribution in galaxy clusters have been limited to the
standard spherical geometry \cite{rb02,Boname06,SHL09}. The importance of the intrinsic geometry of the cluster has been emphasized by many authors \cite{FoxPen02,Jing02,Plionis06,Sereno06} and
the standard spherical geometry has been severely questioned, since Chandra and XMM-Newton observations have shown that clusters usually exhibit an elliptical surface brightness. { In general, different cluster gas profiles give different $\theta_{c}$ values, for instance,  $\theta_{ell} = \frac{2e_{proj}}{1+e_{proj}}\theta_{circ}$ \cite{DeFilippis05}, where  $\theta_{ell}$ and $\theta_{circ}$ are the core radius obtained by using an isothermal elliptical
$\beta$ model and an isothermal spherical $\beta$ model, respectivelly, and $e_{proj}$ is the
axial ratio of the major to the minor axes of the projected isophotes. Therefore, the assumed cluster shape can affect considerably the SZE/X-ray distances, and, consequently, the $H_{0}$ estimates and other astrophysical quantities.}

The first determination of the intrinsic three-dimensional (3D) shapes of galaxy clusters was presented in Ref. \cite{Morandi2010} by combining X-ray, weak-lensing and strong-lensing observations. Their methodology was
applied to the galaxy cluster MACS J1423.8+2404 and it was found the
presence of a triaxial galaxy cluster geometry with DM halo axial
ratios $1.53 \pm 0.15$ and $1.44 \pm 0.07$ on the plane of the sky
and along the line of sight, respectively. { More recently, the elliptical description to  galaxy clusters was shown to be the most compatible with the validity of the cosmic distance duality  relation $D_L (1+z)^ {-2}/D_A=1$, where $D_L(z)$ and $D_A(z)$ are, respectively, the luminosity and angular diameter cosmological distances \cite{eth33,Holandaa,Holandab,Li2011}.}

On the other hand, in 2005, Eisenstein {\it{et al.}}\cite{Eisenstein05} presented the large scale correlation
function from the Sloan Digital Sky Survey (SDSS) showing clear evidence for the baryon
acoustic peak at $100 h^{-1}$ Mpc scale, a result in good agreement with the analyses
from CMB data\cite{Komatsu}. The breaking
on the degeneracy between $H_0$ and $\Omega_m$ by applying this BAO signature is possible
because it does not depend on $H_0$ and is highly sensitive to the matter density parameter.
The combination  SZE/X-ray with BAO has been recently discussed in the literature by Cunha,
Marassi and Lima\cite{cunha}.

In this article, we discuss the determinations of $H_0$ and
$\Omega_m$ by considering the  Bonamente
{\it{et al.}} sample \cite{Boname06}. { This sample is formed by  38 ADD of galaxy clusters at redshifts $0.14 < z < 0.89$, derived from Chandra X-ray and OVRO/BIMA interferometric SZE (see Fig. 1a). In order to perform a realistic model for the cluster gas distribution and take into account the possible presence of the cooling flow, the gas density was modeled  by the non-isothermal spherical double $\beta$-model. This model generalizes the single
$\beta$-model \cite{caval} and it has two core radius that describe the shape of the inner and outer portions of the density
distribution. Analysis of the SZE and X-ray observations has shown that the outer core radius in this model is bigger than ones obtained by using  an isothermal elliptical
$\beta$ model or an isothermal spherical $\beta$ model (see tables in Ref.\cite{Boname06}).} Our basic
findings follow from a joint analysis involving the data from SZE
and X-ray surface brightness with the recent SDSS measurements of
the baryon acoustic peak. The influence of  the intrinsic cluster
geometry  on $H_{0}$ it will be quantified by comparing the results
derived here with a recent determination of both
parameters\cite{cunha} based on De Filippis {\it{et al.}} sample
formed by  25 clusters, where was used a isothermal elliptical $\beta$ model \cite{DeFilippis05}.

\section{Basic Equation and Statistical Analysis}

Let us now consider that the Universe is described by a flat
Friedmann-Robertson-Walker (FRW)  geometry 

\begin{equation}
 ds^2=dt^2-a^2(t)\left[dr^2+r^2(d\theta^2+sin^2\theta d\phi)\right],
\end{equation}
driven by cold dark
matter plus a cosmological constant ($\Lambda$CDM). 

In this background, the angular
diameter distance, ${\cal{D}}_A$, can be written
as\cite{cunha,Alc04}
\begin{equation}
{\cal{D}}_A(z;h,\Omega_m) = \frac{3000h^{-1}}{(1 +
z)}\int_{o}^{z}\frac{dz'}{{\cal{H}}(z';\Omega_m)} \quad
\mbox{Mpc}, \label{eq1}
\end{equation}
where $h=H_0/100$ km s$^{-1}$ Mpc$^{-1}$ and the dimensionless
function ${\cal{H}}(z';\Omega_m)$ is given by ${\cal{H}} = \left[\Omega_m(1 + z')^{3} + (1
-\Omega_m)\right]^{1/2}.$
As it appears, the above expression has only two free parameters ($h,\Omega_{m}$).
In this way, we perform a statistical fit over the $h - \Omega_{m}$
plane in light of Bonamente {\it et al.} data sample\cite{Boname06}.  In our analysis we use a maximum likelihood that can be
determined by a $\chi^2$ statistics,
\begin{equation}
\chi^2(z|\mathbf{p}) = \sum_i { ({\cal{D}}_A(z_i; \mathbf{p})-
{\cal{D}}_{Ao,i})^2 \over \sigma_{{\cal{D}}_{Ao,i}}^2 + \sigma_{stat}^{2}},
\end{equation}
where ${\cal{D}}_{Ao,i}$ is the observational ADD,
$\sigma_{{\cal{D}}_{Ao,i}}$ is the uncertainty in the individual
distance, $\sigma_{stat}$ is the contribution of the statistical errors
(see table 3 in Bonamente {et al.}\cite{Boname06} (2006)) added in quadrature
($\approx 20$\%) and the complete set of parameters is given by
$\mathbf{p} \equiv (h, \Omega_{m})$. { It is interesting to note that for type Ia supernovae  analyses, $h$ is usually marginalized over.}

In what follows, we first  consider the SZE/X-ray distances
separately, and, further, we present a joint analysis including the
BAO signature from the SDSS catalog. Note that a specific flat
$\Lambda$CDM cosmology has not been fixed a priori in the analysis below.

\subsection{Limits from SZE/X-ray}

To begin with, let us consider the influence of the
Hubble parameter on the ADD once the cosmology is fixed.

In  Fig. 1(a), we display the residual
Hubble diagram and the galaxy cluster data  sample by considering
a flat cosmic concordance model ($\Omega_m=0.3,
\Omega_{\Lambda}=0.7$). As expected, for a given redshift, the
distances increase for smaller values of $H_0$.

In Fig. 1(b) we show the contours of constant likelihood (68.3\%,
95.4\% and 99.7\%) in the {{parameter space}} $h-\Omega_m$ for the
SZ/X-ray data above discussed. Note that a large range for the $h$
parameter is allowed, ($0.65 \leq h \leq 0.86$), at $1\sigma$ of
confidence level. In particular, we found
$h=0.755^{+0.065}_{-0.065}$ with $\chi^2_{min}=35.2$ at $68.3$\%
confidence level (c.l.). Naturally, such bounds on $h$ are
reasonably dependent on the cosmological model adopted. For example,
if we fix $\Omega_{m}=0.3$ we have $h=0.76$, for $\Omega_{m}=1.0$ we
have $h=0.65$, and both cases are permitted with high degree of
confidence. This means that using only the Bonamente {\it et al.} sample  we
can not constrain severely the energetic components of the
$\Lambda$CDM model with basis on the SZE/X-ray data alone. 
As one may conclude,  one additional cosmological
test (fixing $\Omega_m$ { or $h$)} is necessary in order to break the degeneracy on the $(\Omega_{m}, h)$ plane. { As we are interested in $H_0$ constraints from galaxy clusters observations, it is interesting to use a cosmological test  sensitive only to $\Omega_{m}$ to break the degeneracy on the $(\Omega_{m}, h)$ plane.}

\begin{figure}[h] 
       \begin{minipage}[b]{0.50 \linewidth}
         {\includegraphics[width=\linewidth]{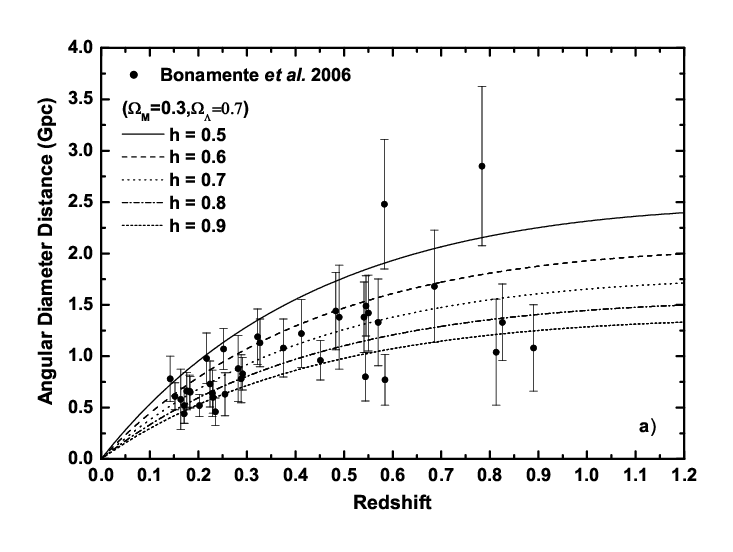}}\\

           \label{fig:XXX}
       \end{minipage}\hfill
       \begin{minipage}[b]{0.50 \linewidth}
           {\includegraphics[width=\linewidth]{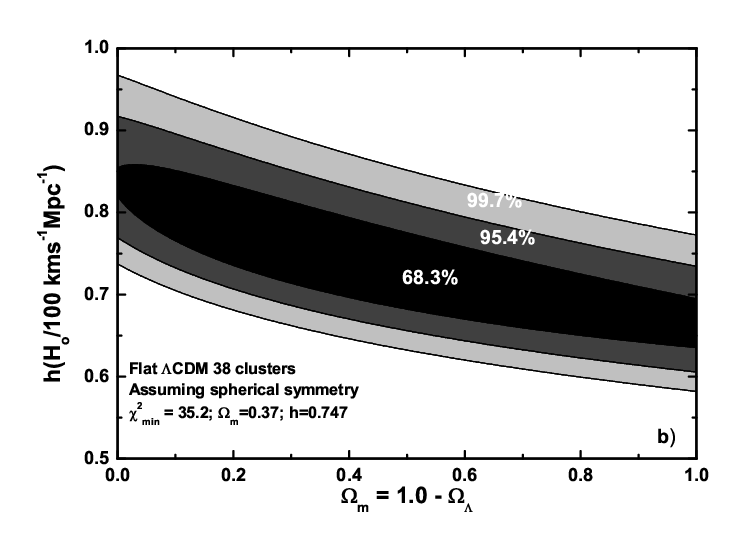}}\\

           \label{fig:XXXX}
       \end{minipage}
       \caption{{\small a) Angular diameter distance as a function of
redshift for $\Omega_{m}=0.3$, $\Omega_{\Lambda}=0.7$ and some
selected values of the $h$ parameter.  The data points correspond to
the the SZE/X-ray distances for 38 clusters from Bonamente {\it et
al.} sample. b) Confidence regions ($68.3$\%, $95.4$\% and
$99.7$\%) in the $(\Omega_{m}, h)$ plane provided by the SZE/X-ray
data. The best fit values are $h = 0.743$ and $\Omega_{m} = 0.37$.
As remarked in the text, the possible values of $H_0$ are heavily dependent on the allowed values of $\Omega_m$,
and, therefore, such a degeneracy need to be broken by adding a new cosmological test.}}
   \end{figure}
\newpage
\subsection{Joint analysis for SZE/X-ray and BAO}

As remarked in the introduction, more stringent constraints on the
{{parameter space}} ($h, \Omega_m$) can be obtained  by combining the
SZE/X-ray with the BAO signature\cite{Eisenstein05}. The peak
detected (from a sample of 46748 luminous red galaxies selected from
the SDSS Main Sample) is predicted to arise  at the
measured scale of 100 $h^{-1}$ Mpc. Let us now consider it as an
additional cosmological test over the spherical cluster sample. Such
a measurement is characterized by

\begin{eqnarray}
 {\cal{A}} \equiv {\Omega_{\rm{m}}^{1/2} \over
 {{\cal{H}}(z_{\rm{*}})}^{1/3}}\left[\frac{1}{z_{\rm{*}}}
 \Gamma(z_*)\right]^{2/3}  = 0.469 \pm 0.017, 
\end{eqnarray}
where $z_{\rm{*}} = 0.35$ is the redshift at which the acoustic
scale has been measured and $\Gamma(z_*)$ is the dimensionless
comoving distance to $z_*$. Note that the above quantity is independent of the Hubble constant,
and, as such, the BAO signature alone constrains only the $\Omega_m$
parameter.

In  Fig. 2(a), we show the confidence regions for the SZE/X-ray
cluster distance and BAO joint analysis. By comparing with Fig.
1(b), one may see how the BAO signature breaks the degeneracy in the
$(\Omega_{\rm{m}}, h)$ plane. As it appears, the BAO test presents a
striking orthogonality centered at $\Omega_m=
0.274^{+0.036}_{-0.026}$ with respect to the angular diameter
distance data  as determined from SZE/X-ray processes. We find $h=
0.765^{+0.035}_{-0.033}$  at $68.3$\% c.l.  and
$\Omega_{m}=0.273^{+0.03}_{-0.02}$ at $68.3$\% c.l.  for $1$ free
parameter ($\chi^2_{min}=35.3$). In light of these results, the important lesson here is that
the combination of SZE/X-ray with BAO provides an interesting
approach to constrain the Hubble constant.

In Fig. 2(b), we have plotted the likelihood function for the $h$
parameter in a flat $\Lambda$CDM universe for the SZE/X-ray + BAO
data set. The dotted lines are cuts in the regions of $68.3$\%
probability and $95.4$\% (1 free parameter).

At this point, it is interesting to compare our results with {{other}}
recent works.  In table 1, a  list of recent Hubble parameter
determinations based on cluster data is displayed. Note that in the
first five works the $h$ values were obtained by fixing the
cosmology ($\Omega_m=0.3$, $\Omega_{\Lambda}=0.7$), while in our
estimate the SZE/X-ray + BAO technique was used (no one specific
cosmology has been fixed).
\begin{figure}[h] 
       \begin{minipage}[b]{0.50 \linewidth}
           {\includegraphics[width=\linewidth]{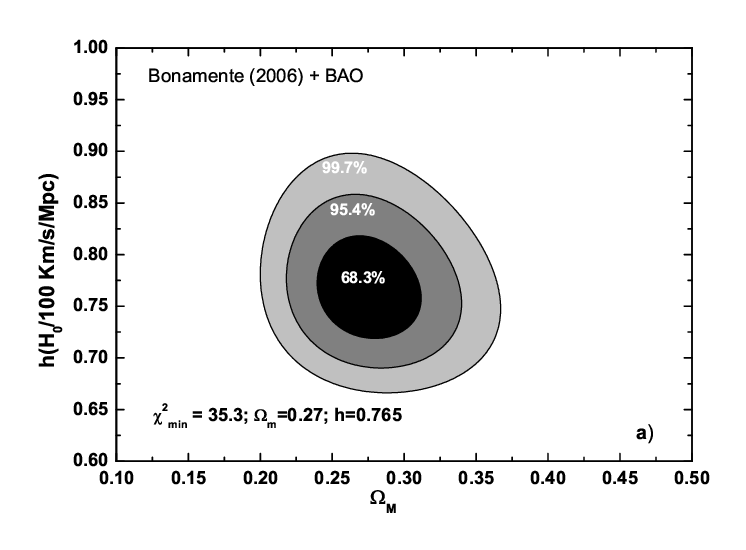}}\\

       \end{minipage}\hfill
       \begin{minipage}[b]{0.50 \linewidth}
            {\includegraphics[width=\linewidth]{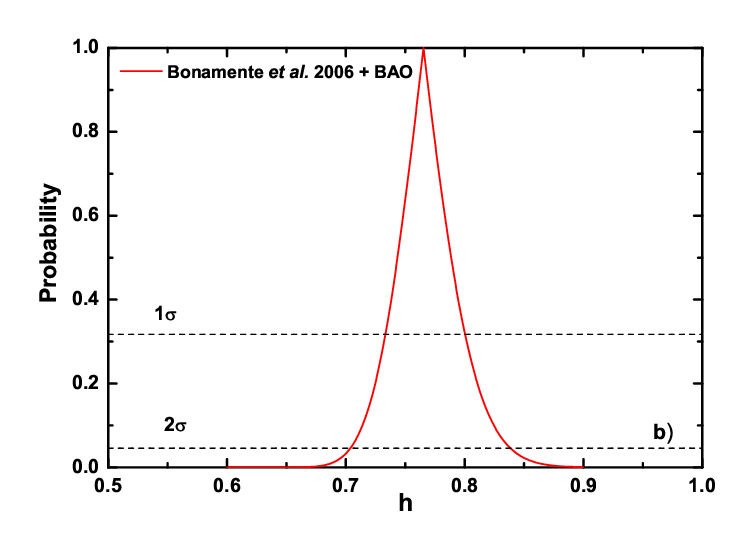}}\\
       \end{minipage}
       \caption{{\small a) Contours in the $\Omega_m - h$ plane using
the SZE/X-ray and BAO joint analysis. The contours correspond to
$68.3$\%, $95.4$\% and $99.7$\% confidence levels. The best-fit
model converges to $h = 0.765$ and $\Omega_m=0.27$. b) Likelihood
function for the $h$ parameter in a flat $\Lambda$CDM universe from
SZE/X-ray emission and BAO. The horizontal lines are cuts in the regions of
$68.3$\% probability and
 $95.4$\%. }}
   \end{figure}
   \begin{table}[ph]
\caption{Limits to $h$ using SZ/X-ray method from galaxy clusters
($\Lambda$CDM)} {\begin{tabular} {@{}ccc@{}}  Reference
(data) & $\Omega_m$ & $h$ ($1\sigma$) \\ \hline \hline
Mason {\it{et al.}} 2001 (7 clusters)\cite{Mas01}  &$0.3$& $0.66^{+0.14}_{-0.11}$ \\
Reese {\it{et al.}} 2002
(18 cluster) \cite{Reese02} &$0.3$&$0.60^{+0.04}_{-0.04}$ \\
Reese 2004 (41 clusters) \cite{Reese04}  &$0.3$&$0.61^{+0.03}_{-0.03}$ \\
Jones {\it{et al.}} 2005 (5 clusters) \cite{Jones05}  &$0.3$&$0.66^{+0.11}_{-0.10}$ \\
Bonamente {\it{et al.}} 2006
(38 clusters) \cite{Boname06}  &$0.3$&$0.77^{+0.04}_{-0.03}$ \\
Cunha {\it{et al.}} 2007 (24 triaxial clusters) +
BAO \cite{cunha}&$0.273^{+0.03}_{-0.02}$&$0.738^{+0.042}_{-0.033}$
\\
{{This paper (38
clusters)+BAO  }}&{\boldmath{$0.273^{+0.03}_{-0.02}$}}&{\boldmath{$0.765^{+0.035}_{-0.033}$}}
\\ \hline
\end{tabular} \label{ta1}}
\end{table}

On the other hand, the importance of the intrinsic geometry of the cluster
has been emphasized by many authors\cite{FoxPen02,Jing02,Plionis06,Sereno06}. 
As a consequence, the standard spherical
geometry has been severely questioned, since Chandra and XMM-Newton
observations have shown that clusters usually exhibit an elliptical
surface brightness. In this concern, a previous determination of $H_0$ from SZE/X-ray + BAO by
Cunha {\it et al.}\cite{cunha} was based in a smaller sample (25
galaxy clusters described by an elliptical profile) observed by De Fillipis and
collaborators\cite{DeFilippis05}. Their results suggested that 15 clusters are in fact more
elongated along the line of sight, while the remaining 10 clusters
are compressed. Actually, the assumed cluster shape seems to affect considerably
the SZE/X-ray distances, and, therefore, the $H_{0}$ estimates.

As shown in table 1, the central value of $H_0$ in this case (elliptical profile) 
is  the same determined by Riess {\it et al.}\cite{Riess11} by
using a hybrid sample (Cepheids + Supernovae), $H_0=73.8 \pm 2.4$ km/s/Mpc . It is also closer to
the Hubble Space Telescope (HST) key project determination of the
Hubble parameter announced by Friedman {\it et al.}\cite{WF01} based only on
Cepheids as distance calibrators,$H_0=72 \pm 8$ km/s/Mpc.

\section{Comments and Conclusions}

In this paper we have discussed a new determination of the Hubble
constant based on the SZE/X-ray distance technique for a sample of
38 clusters as compiled by Bonamente {\it et al.} assuming spherical
symmetry\cite{Boname06}. The degeneracy on the $\Omega_{m}$ parameter was broken
through a joint analysis applying the baryon acoustic oscillation signature from the SDSS
catalog. The Hubble constant was constrained to be $h =
0.765^{+0.035}_{-0.033}$ for $1\sigma$ and
$\Omega_{m}=0.273^{+0.03}_{-0.02}$. These limits were derived
assuming the flat $\Lambda$CDM scenario and a spherical $\beta$-model.
The central $h$ value derived here is in perfect agreement with the Bonamente {\it{et al.}}
value, $h=0.769$ (assuming a hydrostatic equilibrium  and fixing the concordance model), as well as with {{other}} recent estimates coming
from WMAP and Hubble Space telescope Key Project, where $h=0.73$.
However, it differs slightly of a recent estimation  by
Cunha, Marassi \& Lima, $h=0.74$,  where  the SZE/X-Ray +
BAO technique was applied for 24 triaxial galaxies clusters. In general ground, such results
are suggesting that the combination of these three independent phenomena (SZE/X-ray and BAO)
provides an interesting method to constrain the Hubble constant.

{}
\end{document}